%% file: main.tex
\documentclass[
    aps,
    prl,
    twocolumn,
    groupedaddress,
    showpacs,
    floatfix,
    altaffilletter, 
    include graphics, 
    longbibliography,
]{revtex4-2}
\input{preamble}

\begin{document}

    \title{Resilient strange metal at an unconventional quantum critical point in $d=2$}

    \author{Jérôme Leblanc}
    \email{jerome.leblanc@usherbrooke.ca}
    \affiliation{D\'epartement de physique, Regroupement qu\'eb\'ecois sur les mat\'eriaux de pointe $\&$ Institut quantique Universit\'e de Sherbrooke, 2500 Boul. Universit\'e, Sherbrooke, Qu\'ebec J1K2R1, Canada}
    
    \author{A.-M. S. Tremblay}
    \email{andre-marie.tremblay@usherbrooke.ca}
    \affiliation{D\'epartement de physique, Regroupement qu\'eb\'ecois sur les mat\'eriaux de pointe $\&$ Institut quantique Universit\'e de Sherbrooke, 2500 Boul. Universit\'e, Sherbrooke, Qu\'ebec J1K2R1, Canada}
    %

    \date{\today}

    
        \begin{abstract}
        Properties of the two-dimensional Hubbard model with nearest-neighbor hopping are under scrutiny in cold-atom experiments and diagrammatic quantum Monte Carlo.
        Given the controlled nature of these approaches, it is timely to make predictions about strange-metal behavior and its relation to quantum-critical properties.
        Even for interaction strengths below the Mott transition, several peculiarities occur at the quantum critical point separating Fermi liquid and incommensurate spin-density wave order.
        %
        %
        Here, we predict, using the non-perturbative improved two-particle self-consistent approach, that for correlation lengths ranging from about one to one hundred lattice spacings, Kohn anomalies on the underlying Fermi surface lead to unconventional critical exponents. 
        The temperature dependence of the single-particle self-energy acquires strong momentum dependence along the Fermi surface with no clear evidence of Landau quasiparticles.
        Nevertheless, we observe strange-metal behavior, namely, resistivity that scales linearly with temperature as $T\rightarrow{0}$. 
        Our work shows that a linear temperature dependence of resistivity arises without a linear-in-temperature self-energy along the Fermi surface, as is often assumed. 
    \end{abstract}
    \maketitle


\textit{Introduction---}
Understanding phenomena that defy current paradigms can lead to useful insights.
The first phenomenon we are concerned with is the {\it strange metal}.
A strange metal exhibits a resistivity that is linear in temperature over an extended range, down to $T=0$.
It defies the Landau quasiparticle paradigm leading to the $T^2$ resistivity obeyed by simple metals~\cite{Greig_Rowlands_1974}. 
A wide range of materials exhibits strange metal behavior~\cite{yeHoppingFrustrationinducedFlat2024, meierStrangeWayStrangeMetal2024, fang2009, doiron2009, doiron2009, doiron-leyraudLinearTScatteringPairing2010, custersBreakupHeavyElectrons2003, martelli2019, taupin2022, fournierInsulatorMetal1998, mackenzie1996, gurvitch_1987, martin1990, takagi1992, cooper2009, legrosUniversalTlinearResistivity2019,jinLinkSpinFluctuations2011, yuanScalingStrangemetalScattering2022}.
It is distinct from the {\it bad metal} behavior, arising from strong interactions~\cite{pakhira2015, Brown_Mitra_Bakr_2018, Deng_Mravlje_Zitko_Ferrero_Kotliar_Georges}, where resistivity is linear in temperature at large temperature, beyond the Mott-Ioffe-Regel (MIR) limit. 
Strange metal behavior often accompanies quantum critical points (QCPs), which takes us to the second phenomenon that draws our attention.
QCPs~\cite{Lohneysen_Rosch_Vojta_2007} lead to quantum properties observable at unusually large temperatures~\cite{Kopp_Chakravarty_2005, roy2008scaling, Quantum_criticality_2011, Sachdev_book_1999}, contrary to the standard paradigm of phase transitions~\cite{Halperin_Hohenberg_1977}.
This might explain why $T$-linear resistivity not only begins near $T=0$, but also extends to large $T$ in strange metals.

Past studies employed SYK models~\cite{chowdhurySYK2022, sachdevgapless1993, varmaphenomenology1989, parcolletNonFermi1999, geoargesQuantumFluctuations2001, liStrangeMetal2024} and holographic duality \cite{hartnollStrangeMetallicHolography2010, hartnollScalingTheory2015, hartnollTheoryUniversalIncoherent2015} to investigate the non-Fermi-liquid regime and linear resistivity.
Here, we study the two-dimensional Hubbard model on a square lattice with only nearest-neighbor hopping.
A QCP occurs in this model at finite doping when a $T=0$ long-range ordered incommensurate spin-density wave (SDW)~\cite{Schulz_1990}~\footnote{There has been a claim that instead of being continuous, the transition could be first-order~\cite{Altshuler_Ioffe_Millis_1995}. Our approach allows this possibility, but would require extensive work.} disappears to be replaced at larger doping by a correlated Fermi liquid. 
At small doping and finite temperature, one finds a pseudogap~\cite{Vilk_Tremblay_1995, vilk_temblay_1997, vilk2024antiferromagnetic, Kyung2004, kyung2006, senechal_2004_pseudogap,ferreroPseudogap2009,
gullMillis2013,
fratinoPseudogap2016,
huscroft2001,
macridinpseudogap2006,
stanescuFermiArcs2006,
sakaiEvolution2009,
sordiStrongCoupling2012,
sordiCAxis2013,
gullMillisLetters2013,
reymbautPseudogap2019,
dahsPseudogap2019,
walshPrediction2022,
sordiSpecificHeat2019,
walshCritical2019,
sordiPseudogapTemperatureWidom2012}.
The improved two-particle self-consistent approach that was introduced and benchmarked in Ref.~\cite{gauvin-ndiaye_improved_2023} captures this physics. 
This non-perturbative method satisfies several exact sum-rules, as well as the Mermin-Wagner theorem in two dimensions. Namely, it finds an ordered state only at zero temperature and can be applied in the pseudogap regime asymptotically close to $T=0$~\cite{ vilk2024antiferromagnetic}.
Recent calculations give excellent agreement~\cite{vilk2026pseudogapfermiliquidvan} with cold-atom experimental results~\cite{Kendrick_Kale_Gang_Deters_Lebrat_Young_Greiner_2025,chalopin2025} for the compressibility and Knight shift.

Studying the two-dimensional Hubbard model on a square lattice with only nearest-neighbor hopping is timely because of advances in cold-atom quantum simulators~\cite{Kendrick_Kale_Gang_Deters_Lebrat_Young_Greiner_2025, brown_2019, brown2020, anderson2019, chalopin2025, cocchi2016, pasqualetti2024, richard2025} and diagrammatic quantum Monte Carlo (DiagQMC)~\cite{Simkovic_Rossi_Georges_Ferrero_2024}.
While a finite value of second-neighbor hopping $t'$ is necessary for realistic materials, our choice $t'=0$ allows future direct comparisons with cold atoms and DiagQMC. 
Particle-hole symmetry makes the results identical for positive and negative deviations from half-filling $n=1$.
We study fillings larger than unity. 

The canonical treatment of the antiferromagnetic QCP in $d=2$ is the Hertz~\cite{Hertz_1976} and Millis~\cite{millis1993effect} theory. 
The self-consistent theory of Moriya~\cite{ moriya1990antiferromagnetic} exhibits several similarities. 
Microscopic parameters in these theories are temperature independent.

Here, we demonstrate that:

i) The QCP exhibits unconventional behavior: modulo logarithmic corrections, the set of critical exponents $\gamma$, $\nu$ and $z$ is different from that found in previous studies~\cite{Hertz_1976, millis1993effect}. 
This is because the SDW wave vector $\mathbf{Q}$ connects points of the Fermi surface, so-called hot spots, with nearly antiparallel Fermi velocities. 
These points are ``Kohn points''~\cite{Holder_Metzner_2014} that lead to Kohn anomalies~\cite{Kohn_1959, Stern_1967}.
The Fermi velocities at the hot spots of the renormalized Fermi surface are closer to being antiparallel than for the starting non-interacting Fermi surface (see End Matter). 
%
%
Prior work discusses the robustness to interactions of Kohn points in $d=3$~\cite{Schafer_Katanin_Held_Toschi_2017}.

ii) The unconventional critical behavior follows not only from Kohn points~\cite{Schafer_Katanin_Held_Toschi_2017} but also from the near-nesting of the Fermi surfaces connected by $\mathbf{Q}$. 
This leads to temperature-dependent microscopic parameters, such as Landau damping.
Refs.~\cite{Dare1996Crossover, Bergeron2012QCP, Vilk_Van_Hove_2023} studied different cases of near-nesting that lead to different temperature dependencies.
Earlier studies have examined the $T=0$ consequences for the frequency dependence at the hot spots in different models~\cite{Holder_Metzner_2014, Sykora_Holder_Metzner_2018, Debbeler_Metzner_2023, Debbeler_Metzner_2024, Fus}.
iii) The temperature dependence of the resistivity is linear at the QCP, even though the temperature dependence of the $\omega=0$ self-energy is a strong function of momentum.
This challenges the usual assumption that strange metal behavior implies $T$-linear dependence of the self-energy. 
Landau quasiparticles are absent in the range of $T$ accessible.
Earlier works~\cite{Holder_Metzner_2014, Wang_Chubukov_2013} focus on the $T=0$ frequency dependence at the incommensurate SDW spanning the Fermi surface.

\paragraph{Model---} 
The Hubbard Hamiltonian on a square lattice is
\begin{equation}
    \mathrm H
    =
    \sum_{ij\sigma} t_{ij} \mathrm c^\dag_{i\sigma} \mathrm c _{j\sigma}
    +
    U \sum_i \mathrm n_{i\uparrow} \mathrm n_{i\downarrow},\label{eq:hamiltonian_peierls}
\end{equation}
where $t_{ij}\equiv -t$ are hopping amplitudes between nearest-neighbor sites $i$ and $j$, $\mathrm c^{(\dagger)}_{i\sigma}$ are fermionic operators that remove (create) an electron of spin $\sigma$ on site $i$, $\mathrm n_{i\sigma}\equiv \mathrm{c}^\dagger_{i\sigma}\mathrm{c}_{i\sigma}$ counts the number of electrons of spin $\sigma$ on site $i$ and $U=6t$ is the same-site screened Coulomb repulsion.
 Further-neighbor hoppings $t^\prime,t^{\prime\prime}$ are zero.
The lattice spacing, the electron charge, as well as $\hbar$  and $k_\mathrm{B}$ are set to unity. 
We restrict ourselves to nearest-neighbor hopping where benchmarking~\cite{gauvin-ndiaye_improved_2023} of TPSC+ for charge and spin susceptibilities is available. 
\paragraph{Method---}
We use the Improved two-particle self-consistent approach (TPSC+)~\cite{gauvin-ndiaye_improved_2023, vilk2024antiferromagnetic}, which is an extension of TPSC \cite{vilk_temblay_1997}.
Both are non-perturbative methods for the one-band Hubbard model. They are valid from weak to intermediate interaction strength ($U\lesssim 6 t$).
They also respect the Mermin-Wagner theorem and the Pauli principle.
%
TPSC+ has the advantage of being valid deeper in the renormalized classical regime \cite{vilk2024antiferromagnetic} and of having better benchmarks than TPSC for the magnetic correlation length at low temperatures \cite{gauvin-ndiaye_improved_2023, Schafer_Wentzell_et_al._2021}.
This method satisfies the \textit{RPA-like} sum rules for spin and charge susceptibilities. 
The latter are defined respectively as
\begin{align}
    \chi_\mathrm{sp}(\mathbf{q}, iq_n)^{-1} = \chi^{(2)}(\mathbf{q}, iq_n)^{-1}-U_\mathrm{sp}/{2},\label{eq:chisp2}\\
    \chi_\mathrm{ch}(\mathbf{q}, iq_n)^{-1} = \chi^{(2)}(\mathbf{q}, iq_n)^{-1}+U_\mathrm{ch}/{2},
\end{align}
where $U_\mathrm{sp/ ch}$ are the irreducible spin and charge vertices that we find self-consistently using the spin and charge sum rules 
alongside the TPSC \textit{ansatz}
\begin{equation}
    U_\mathrm{sp}=U
    \frac{\left<\mathrm{n}_\uparrow \mathrm{n}_\downarrow\right>}{\left<\mathrm{n}_\uparrow \right>\left< \mathrm{n}_\downarrow\right>}
    \label{eq:ansatz}.
\end{equation}
The irreducible particle-hole susceptibility is 
\begin{multline}
    \chi^{(2)}(\mathbf{q}, iq_n)\notag\\
    =-\frac{T}{N}\sum_{\mathbf{k}, ik_n}\left[ G^{(2)}(\mathbf{k}, ik_n)G^{(1)}(\mathbf{k}+\mathbf{q},ik_n+iq_n) \right.\notag\\
    +\left. G^{(2)}(\mathbf{k}, ik_n)G^{(1)}(\mathbf{k}-\mathbf{q},ik_n-iq_n) \right]
\end{multline}
with $T$ the temperature and $N$ the number of lattice sites.
In the above expressions, $\mathbf{k}$ and $\mathbf{q}$ are wave vectors, $ik_n$ and $iq_n$ are fermionic and bosonic Matsubara frequencies, $G^{(1)}$ is the non-interacting ($1^\mathrm{rst}$ level approximation) Green's function and $G^{(2)}$ is the interacting ($2^\mathrm{nd}$ level approximation) Green's function. 
The latter includes feedback from the self-energy
\begin{multline}
    \Sigma^{(2)}(\mathbf{k}, ik_n) = U \frac{n}{2} + \frac{T}{N}\frac{U}{8}\sum_{\mathbf{q}, iq_n}\left[ 3U_\mathrm{sp}\chi_\mathrm{sp}(\mathbf{q}, iq_n)\right.\notag\\
    + \left. U_\mathrm{ch}\chi_\mathrm{ch}(\mathbf{q}, iq_n)\right] G^{(1)}(\mathbf{k}+\mathbf{q}, ik_n+iq_n),
\end{multline}
%
%
See Ref.~\cite{gauvin-ndiaye_improved_2023} for further details of this approach.
\paragraph{Spin fluctuations at the QCP---}
For small dopings, the ground state is a SDW. 
The QCP occurs at the doping where the SDW disappears.
It is characterized by a power-law behavior of the zero-frequency spin susceptibility extending to $T=0$.
From now on, we neglect possible logarithmic dependencies on temperature.

\begin{figure}[h]
    \centering
    \includegraphics[width=1\linewidth]{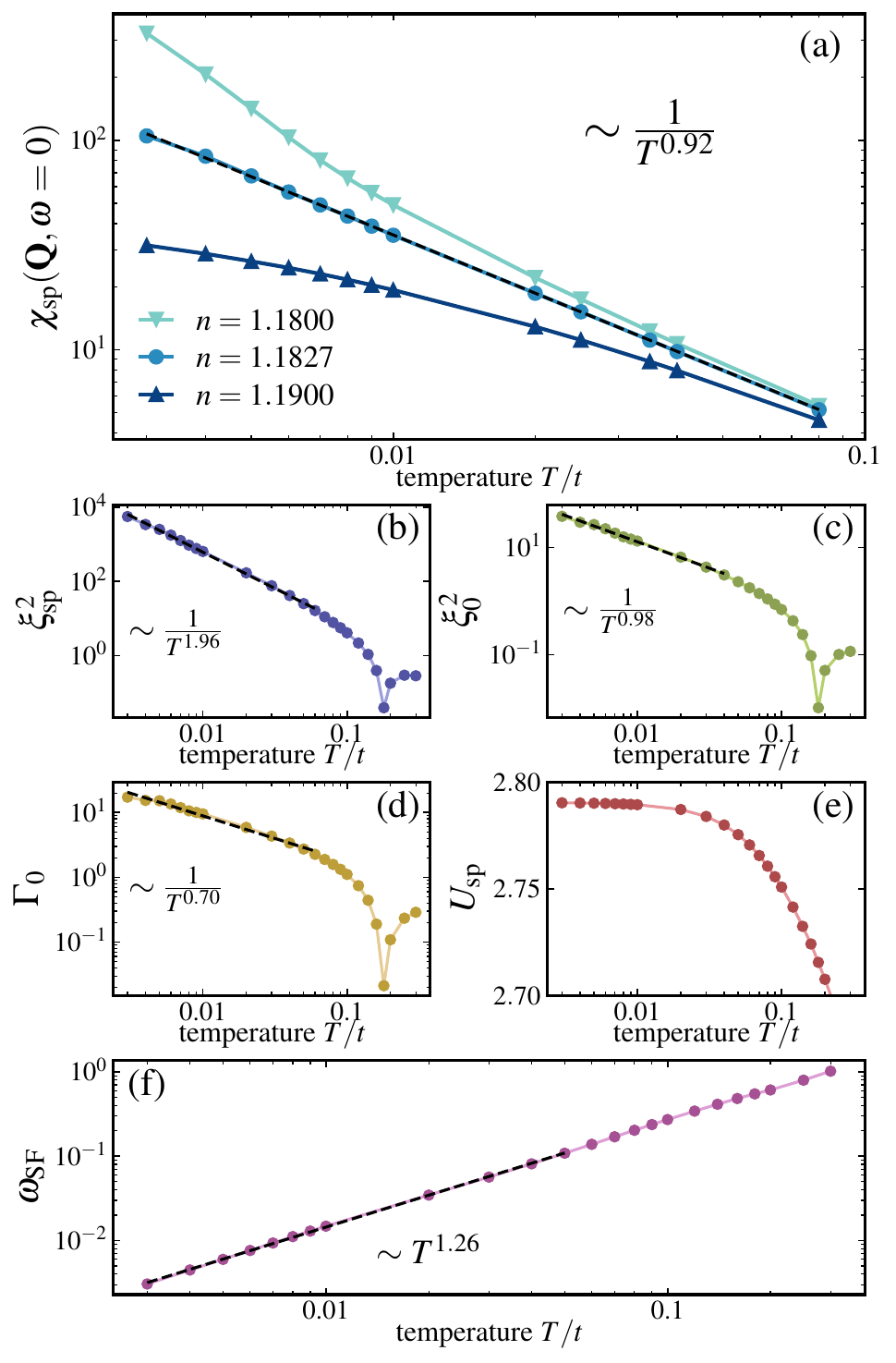}
    \caption{
    Temperature dependence of spin fluctuation-related quantities at the QCP ($n=1.1827$, circles) for $t'=t''=0$, $U=6t$ and at $\mathbf{Q}=(\pi-\epsilon, \pi)$.
    Black dashed lines are power law fits at low temperatures.
    The resulting exponents are indicated by the insets $\sim T^a$.
    The estimated error on $a$ are $\pm0.003$ for (a), $\pm 0.02$ for (b-d) and $\pm0.01$ for (f).
    (a) Maximum of spin susceptibility at zero frequency.
    (b) Squared spin correlation length.
    (c) Squared microscopic particle-hole coherence length scale. Obtained from the second derivative with respect to $q_x$.
    (d) Landau damping constant.
    (e) Irreducible spin vertex.
    (f) Characteristic spin fluctuation frequency obtained from $\Gamma_0/\xi_\mathrm{sp}^{2}$.
    }
    \label{fig:magnum_opus}
\end{figure}

The critical doping follows from the log-log plot in \figref{fig:magnum_opus}(a) for $\chi_\mathrm{sp}(\mathbf{Q}_i,0)$ (\equref{eq:chisp2}) as a function of $T$ at the wave vector $\mathbf{q}=\mathbf{Q}_i$ where it is maximal. 
We find a $1/T^{\gamma}$  temperature dependence with $\gamma\sim0.92$ for filling $n=1.1827$. 
Deviations from this trend at other dopings confirm that this is a good approximation for the quantum-critical doping.
The value of $\mathbf{Q}$ moves from incommensurate $\mathbf{Q}_i$ to $(\pi,\pi)$ for $T>0.1t$. 

As shown in the supplementary material (SM)~\cite{SM}, $\chi_\mathrm{sp}(\mathbf{q}-\mathbf{Q}_i,0)$ has a parabolic dependence on wave vector for small $\mathbf{q}-\mathbf{Q}_i$.
It is maximal at four wave vectors, $\mathbf{Q}_i$, symmetrically located along the zone boundary at $(\pi\pm\epsilon,\pi)$ and $(\pi,\pi\pm\epsilon)$ in the asymptotic low-temperature limit. 
We thus assume an Ornstein-Zernicke form for the retarded $\chi_\mathrm{sp}$~\cite{Dare1996Crossover}
\begin{multline}\label{eq:chi_OZ}
    \chi_\mathrm{sp}(\mathbf{q}, \omega) \\ \approx \frac{2}{U_\mathrm{sp}}\frac{\xi_\mathrm{sp}^2}{\xi^2_0}\sum_i \left( \frac{1}{1+(\mathbf{q}-\mathbf{Q}_i)^2\xi_\mathrm{sp}^2 - i\omega\xi_\mathrm{sp}^2/\Gamma_0} \right),
\end{multline}
with the microscopic particle-hole coherence length scale 
\begin{equation}
    \xi_{0}^2 \equiv -\frac{1}{2\chi^{(2)}(\mathbf{Q}_i, 0)}
    \left.
    \frac{\partial^2\chi^{(2)}(\mathbf{q}, 0)}{\partial q_{x(y)}^2}
    \right|_{\mathbf{q}=\mathbf{Q}_i},\label{eq:xi02}
\end{equation}
the microscopic Landau damping constant $\Gamma_0$ 
\begin{equation}
    \frac{1}{\Gamma_0} \equiv \frac{1}{\xi^2_0}\frac{1}{\chi^{(2)}(\mathbf{Q}_i, 0)}\left.\frac{\partial \mathrm{Im}~\chi^{(2)}(\mathbf{Q}_i, \omega)}{\partial \omega}\right|_{\omega=0},
    \label{eq:gamma_0}
\end{equation}
and the SDW correlation length
\begin{equation}
    \xi_\mathrm{sp}^2
    \equiv
    \xi^2_0\left(
        \frac{2}{U_\mathrm{sp}\chi^{(2)}(\mathbf{Q}_i, 0)}-1    
    \right)^{-1}.\label{eq:xi_cp^2}
\end{equation}
In the Ornstein-Zernicke form, $\omega_\mathrm{SF}=\Gamma_0/\xi_\mathrm{sp}^2$ is the characteristic spin fluctuation frequency. 
We have verified that values of second derivatives along $q_x$ and $q_y$ in \equref{eq:xi02} are close to each other, if not identical, at the QCP (see SM~\cite{SM}) even though they could, in principle, differ in the incommensurate case.

We can extract the temperature dependence of the correlation length $\xi_\mathrm{sp}^2$ directly from the fit of \equref{eq:xi_cp^2}
in Fig.~\ref{fig:magnum_opus}(b)
if we know the temperature dependence of $\xi_{0}^2$ and $U_\mathrm{sp}$.
When $t'\ne0$, $\xi_{0}^2$ is asymptotically independent of $T$, as in the usual Hertz-Millis theory~\cite{millis1993effect}.
As discussed in the End Matter, Kohn anomalies lead to a non-trivial $\xi_{0}^2\sim 1/T^{0.98}$ dependence, as illustrated in Fig.~\ref{fig:magnum_opus}(c). 
The downturn around $T\sim0.2t$ in that figure and also in parts (b) and (d) arises from a shift of the maximum of the susceptibility from $(\pi,\pi)$ to incommensurate wave vectors. 
We deduce that $\xi_\mathrm{sp}^2 \sim \chi_\mathrm{sp}(\mathbf{Q}_i,0) \xi_{0}^2\sim 1/T^{0.92+0.98}\sim 1/T^{1.90}$, which is consistent with~\figref{fig:magnum_opus}(b) where $\xi_\mathrm{sp}^2\sim1/T^{
2\nu}\sim 1/T^{1.96}
$, or $\nu=0.98$, obtained from the definition involving $U_\mathrm{sp}$ in~\equref{eq:xi_cp^2} above.
While $U_\mathrm{sp}$ in general depends on temperature in TPSC,~\figref{fig:magnum_opus}(e) shows that this temperature dependence becomes asymptotically negligible.

The dynamical critical exponent $z$ is the last one we need at a QCP.
Given $\xi_\mathrm{sp}^2\sim 1/T^{1.96}$, the characteristic spin fluctuation frequency 
in \figref{fig:magnum_opus}(f) follows from the temperature dependence obtained in \figref{fig:magnum_opus}(d) for the Landau damping $\Gamma_0\sim 1/T^{0.70}$ in \equref{eq:gamma_0}. 
Indeed, $\omega_\mathrm{SF}$
scales as $T^{1.26}$ since
$\omega_\mathrm{SF}=\Gamma_0/\xi_\mathrm{sp}^2\sim T^{1.96}/T^{0.7}$.
See \cite{SM} that shows why the downturn is absent in \figref{fig:magnum_opus}.
From $\omega_\mathrm{SF}\sim T^{z\nu}$ we find the dynamical exponent $z=1.26/0.98=1.28$.

To summarize, with $\gamma=0.92$, $\nu=0.98$ and $z=1.28$ we can write the scaling relation 
\begin{multline}\label{eq:scaling_fct}
    \chi_\mathrm{sp}(T,\mathbf{q-Q_i},\omega)- R = 
s^{\gamma/\nu}\chi_\mathrm{u}(s^{1/\nu} T, s(\mathbf{q-Q_i}), s^z\omega)\nonumber
\end{multline}
where $R$ is a background and $\chi_\mathrm{u}$ is normally a universal function of its arguments. Here however, the critical behavior is unconventional and specific to the Hubbard model with nearest-neighbor hopping.
These exponents do not correspond to any known universality class. 
When $t'\ne 0$ the exact critical exponents in $d=2$~\cite{Schlief_Lunts_Lee_2017}, are $\nu=1$ and $z=1$.
We are below the upper critical dimension since $d+z<4$.


%
%
%

One can determine the spin-correlation length self-consistently from the local spin sum-rule.
For an Ornstein-Zernike susceptibility, that can be done analytically, as in Ref.~\cite{roy2008scaling}.
It involves logarithmic corrections. 
However, the results do not coincide with what we find because, despite a quadratic maximum in $\mathbf{q}$ in the susceptibility, the integral involves all wave vectors. 
The Ornstein-Zernicke fails rapidly at large wave vectors in two dimensions because in the zero-temperature limit, the susceptibility exhibits non-analytic square-root cusps that were attributed to ``pseudo-nesting'' ~\cite{Gabay_Beal-Monod_1978, Benard_Chen_Tremblay_1993, Sykora_Holder_Metzner_2018}.

\begin{figure}[h]
    \centering
    \includegraphics[width=0.8\linewidth]{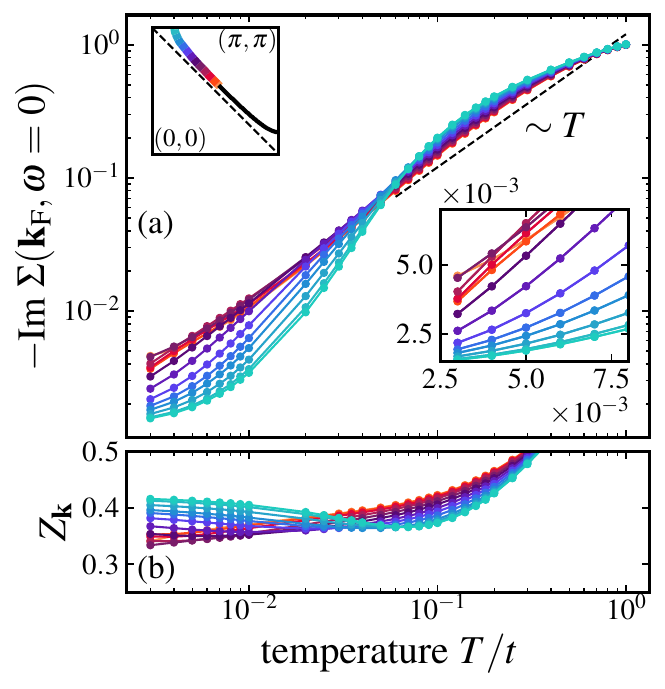}
    \caption{
    (a) Temperature dependence of the imaginary part of the self-energy along the Fermi surface
    at the QCP ($n=1.1827$) for $t'=t''=0$, $U=6t$. 
    The dashed line is a guide to the eye.
    %
    %
    The inset in the upper-left corner displays the color code.
    The full black line is the maximum of the Fermi surface, while the diagonal dashed line is the AFM Brillouin zone.
    The inset on the lower right shows values at low temperature near the node. This inset has linear axes.
    (b) Temperature dependence of the single-particle weight along the Fermi surface for the same parameters. 
    Color code in the inset of (a).
    }
    \label{fig:scaling_self_omega_0_fermi_surface}
\end{figure}
\paragraph{Self-energy---}

\figref{fig:scaling_self_omega_0_fermi_surface}(a) shows the temperature dependence of the imaginary part of the TPSC+ self-energy at the QCP.
At low temperatures, we see a clear distinction between the node and the anti-node. 

No parts of the self-energy along the Fermi surface exhibit a clear power-law dependence on temperature over an extended range.
The inset of \figref{fig:scaling_self_omega_0_fermi_surface}(a) shows that, as we get closer to the node from the anti-node, the self-energy starts off increasing, then decreases as it approaches $\mathbf{k} = (\pi/2,\pi/2)$.
Additionally, it changes curvature as we reach the node.
Briefly, this self-energy has no uniform temperature dependence along the Fermi surface.

At high temperatures, the self-energy at the node tends toward a linear dependence, but over a very narrow range. 
A saturation over the whole Fermi surface follows, which is the first sign of reaching the MIR limit.
Indeed, because of our choice of units, we reach this limit when $\ell \sim v_\mathrm{F}\tau \sim 1$, where $\ell$ is the mean free path, $v_\mathrm{F}$ is the Fermi velocity, and $\tau^{-1}\sim\mathrm{Im}~\Sigma$ is the electron lifetime. 

Unlike earlier work~\cite{Bergeron2012QCP}, the behaviour of the self-energy cannot be linked to the Ornstein-Zernike form of the susceptibility. In particular, factorization of the $\omega/T$ dependence of the Bose and Fermi factors is impossible because $\Gamma_0$ scales with an exponent less than unity. See SM~\cite{SM} for more.
The single-particle weight in \figref{fig:scaling_self_omega_0_fermi_surface}(b) is obtained from Matsubara frequencies $Z_\mathbf{k}^{-1}=\lim_{k_n\to 0}(1-\mathrm{Im}~\Sigma(\mathbf{k}, ik_n)/k_n)$.
Ref.~\cite{gauvin_2022_resilient_fermi_liquid} shows that even in the presence of a pseudogap, we still have well-defined quasiparticles near $(\pi, 0)$ when $t'\ne 0$, namely $T^2$ behavior and a temperature-independent $Z_\mathbf{k}$.  
At the $t'=0$ QCP, there are important differences with that case. 
\figref{fig:scaling_self_omega_0_fermi_surface}(b) shows $Z_\mathbf{k}$ being strongly dependent on temperature and on wave-vector, also in sharp contrast with Ref.~\cite{Deng_Mravlje_Zitko_Ferrero_Kotliar_Georges}.
Even at the anti-node, then, the quasiparticle picture fails. 
However, given the tendency to saturation, it may reappear close to $(\pi,0)$ at a lower temperature. 

There is a small region where $Z_\mathbf{k}$ is independent of temperature, but this is not sufficient to claim that there is a Fermi liquid at this wave vector since the temperature dependence of $\rm{Im}\Sigma$ differs from $T^2$.
There are no well-defined quasiparticles in the Landau sense.
\begin{figure}[h]
    \centering
    \includegraphics[width=0.8\linewidth]{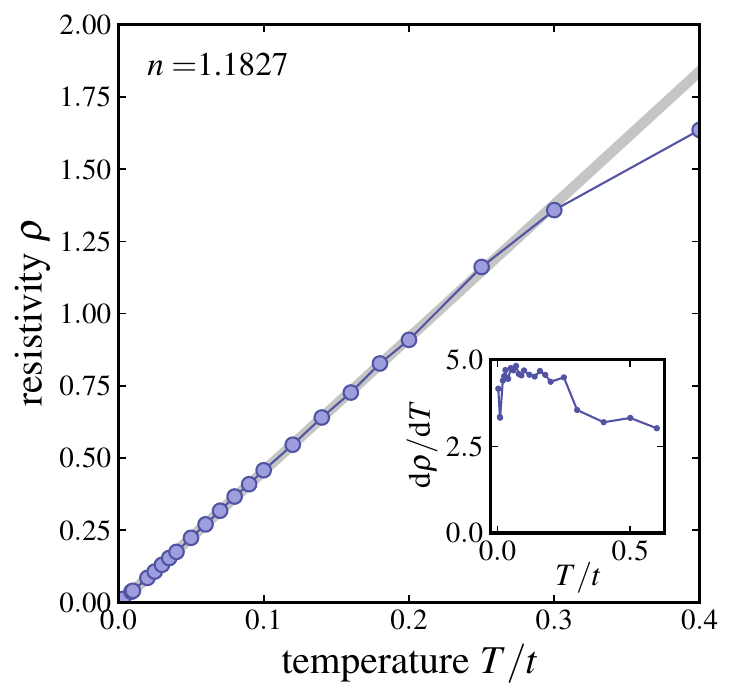}
    \caption{
    Temperature dependence of resistivity at the SDW QCP ($n=1.1827$) for $t'=t''=0$, $U=6t$.
    Blue dots are values of resistivity, and the gray line is a linear fit for $T\leq 0.2t$.
    The inset shows the derivative with respect to temperature.
    }
    \label{fig:rho-GG}
\end{figure}

\paragraph{Resistivity---}
The resistivity follows from the Kubo expression for longitudinal conductivity, i.e.,\label{eq:re_sigma} $\mathrm{Re}~\sigma_{xx}(\omega) = \mathrm{Im}~\chi_{\mathrm{j}_x\mathrm{j}_x}(\omega)/{\omega}$
with $v_x=\partial\epsilon_\mathbf{k}/\partial k_x$ and
\begin{equation}
    \chi_{\mathrm{j}_x\mathrm{j}_x}(iq_n)
    \notag
    =
    -\frac{2T}{N}\sum_{\mathbf{k},ik_n}
    v_x^2
    G^{(2)}(\mathbf{k}, ik_n)
    G^{(2)}(\mathbf{k}, ik_n+iq_n)
    \notag\\
    \label{eq:bubble_chijj_iqn}
\end{equation}
the $\mathbf{q}=\mathbf{0}$ current-current response function where
$\mathrm{j}_x$
is the $x$ component of the current operator $\mathbf{j}$ \cite{bergeron_optical_2011,arsenault_transport_2013, paul2003, reymbaut_maximum_2017}.
We neglect vertex corrections.
Prior work accounting for them in TPSC~\cite{bergeron_optical_2011} satisfies the $f$-sum rule (See End Matter) exactly.
They find that vertex corrections only change the slope of the linear dependence on temperature of the resistivity, not the linearity itself. 
The End Matter discusses that sum rule. 
We use the shorthand $\rho\equiv 1/(\mathrm{Re}~\sigma_{xx}(0)$) for dc resistivity and perform analytical continuation using maximum entropy~\cite{bergeron_algorithms_2016}.

\figref{fig:rho-GG} shows the resistivity as a function of temperature at the QCP.
Despite the peculiar power laws found for the microscopic quantities and the temperature dependence of the self-energy, the resistivity remains linear for a significant range of temperature, from the lowest temperature, $T= 0.003t$, to $T\sim 0.3t$.
Its slope is slightly larger than that corresponding to Planckian dissipation with $\alpha\sim 2$ (discussed in SM~\cite{SM}).
The resistivity and its derivative in the inset of \figref{fig:rho-GG} show a saturation at higher temperatures which is indicative of the MIR limit.
This limit is expected for the $U$ considered, which is below the Mott transition at half-filling. 
This is consistent with the self-energy in \figref{fig:scaling_self_omega_0_fermi_surface}(a) that saturates at high temperatures. See the SM~\cite{SM} for insights on the difficulties associated with analytical derivations of the power laws.
%

%
%


%
\paragraph{Discussion and conclusion---}
We found strange metal behavior with almost Planckian behavior ($1/\tau_\mathrm{transport}\sim 2T)$ in our non-perturbative calculation of resistivity at the QCP of the one-band nearest-neighbor Hubbard model on the square lattice, despite the unconventional exponents at the QCP and despite the value of the interaction being less than the bandwidth.
Landau quasiparticles are absent at this QCP.
The linear dependence of the resistivity on temperature is thus a robust result. 
The origin of this linear $T$ is different from that originating from phonons.
While the characteristic spin-fluctuation frequency vanishes at $T=0$, it never saturates, which it must in the case of phonons to find linear $T$ from the limiting $T/\omega_\mathrm{D}$ behavior of the Bose factor for $T>\omega_\mathrm{D}$. 

The strange metal behavior found previously in Ref.~\cite{bergeron_optical_2011} originates from a different unconventional QCP with commensurate $\mathbf{Q}$.
It nevertheless shares many similarities with our case, namely Kohn points and temperature dependence of microscopic parameters. 

Future work should investigate vertex corrections~\cite{Bergeron2012QCP}, disorder~\cite{Rosch_1999, Petel_Sachdev_Disorder} and resilient Fermi liquid~\cite{gauvin_2022_resilient_fermi_liquid,hlubinaresistivity1995} effects at asymptotically low temperature ($T\ll 0.003t$).

\textit{Acknowledgments---}
We are grateful to 
A. Chubukov, A. Georges, A.Gleis, A. Toschi, C. Duffy, C. Gauvin-Ndiaye, C. Lahaie, C. Iorio Duval, G. Grissonnanche, J.P.F Leblanc, M. Ferrero, N. Hussey, P. Coleman, Q. Si, S. Verret, and Y. Vilk for insightful discussions.
We are also grateful to D. Bergeron for help regarding the analytic continuation software.
We are indebted to L.-T. Gendron and to S. Verret for careful reading of the manuscript. S. Verret also contributed key ideas. 
The initial version of the program used to perform the calculations was written by C. Gauvin-Ndiaye, C. Lahaie and Y. M. Vilk.
This work has been supported by the Natural Sciences and Engineering Research Council of Canada (NSERC) under Grant No. RGPIN-2019-05312 and RGPIN-2024-05206 (A.-M.S. T.), Canada Graduate Scholarships-
Master’s Program, the Fonds de Recherche du Québec
Nature et Technologies under the  \href{https://doi.org/10.69777/345545}{master's research scholarships program}  (J. L.). 
Simulations were performed on computers provided by the Canadian Foundation for Innovation, the Minist\`ere de l'\'Education des Loisirs et du Sport (Qu\'ebec), Calcul Qu\'ebec, and the Digital Research Alliance of Canada. All authors benefit from their \href{https://doi.org/10.69777/309032}{RQMP membership}. We acknowledge the support of the Natural Sciences and Engineering Research Council of Canada (NSERC), NSERC CREATE/ 575280-2023 - Training in Materials for Quantum Technologies (MaQTech).
\paragraph{Data availability---}
The data that support the findings of this article are publicly available [REFERENCE].
The custom scripts used in the analysis are also publicly available [REFERENCE], but the software package that produced said data is not yet available since further optimizations are being added.
$\Omega$MaxEnt \cite{Bergeron_OmegaMaxEnt_Github} was used to perform analytic continuation of the current response function in this article.
\bibliography{bibli.bib}
\appendix

\section*{End Matter}

The end matter provides details on gauge invariance, $f$-sum rule and Kohn anomalies.

\paragraph{Appendix: Gauge invariance and $f$-sum rule---}

To represent the uniform electric field, we use a gauge in which the scalar potential vanishes, and the electric field arises solely from the vector potential. 
Using
\begin{widetext}
\begin{equation}
    \delta \left\langle \mathrm{j}_{x}^{A}(q_{x}=0,\omega )\right\rangle 
    =
    \left[
\chi _{\mathrm{j}_{x}\mathrm{j}_{x}}(q_{x}=0,\omega )-\chi _{\mathrm{j}_{x}\mathrm{j}_{x}}(q_{x}=0,\omega=0 )\right]
A_{x}(q_{x}=0,\omega )
\label{eq:response_to_A}
\end{equation}
for the response of the retarded current to the vector potential,
guarantees that the response to time-independent (zero frequency) vector potential vanishes.
Adding a constant to the vector potential is an allowed residual gauge transformation.
The above expression thus ensures that the physics is invariant under this gauge transformation.

Dropping $q_x$ and writing the electric field as the time derivative of the vector potential, the expression for the conductivity in terms of the imaginary part of the retarded current-current correlation function then reads


\begin{align}
\sigma _{xx}(\omega )& =\frac{1}{i(\omega +i\eta )}\left[ \int \frac{%
d\omega'}{\pi }\frac{\chi''_{\mathrm{j}_x\mathrm{j}_x}(\omega ^{\prime })}{\omega'-\omega -i\eta }-\int \frac{%
d\omega ^{\prime }}{\pi }\frac{\chi''_{\mathrm{j}_x\mathrm{j}_x}(\omega ^{\prime })}{\omega ^{\prime }}\right]
\label{OpticalCondVectorPotential} \\
& =\frac{1}{i(\omega +i\eta )}\left[ \int \frac{d\omega ^{\prime }}{\pi }%
\frac{\chi''_{\mathrm{j}_x\mathrm{j}_x}(\omega ^{\prime })(\omega
+i\eta )}{\omega ^{\prime }\left( \omega ^{\prime }-\omega -i\eta \right) }%
\right],
\end{align}%
\end{widetext}
so that the conductivity becomes
\begin{equation}
\sigma _{xx}(\omega )=\frac{1}{i}\left[ \int \frac{d\omega'}{\pi }\frac{\chi''_{\mathrm{j}_x\mathrm{j}_x}(\omega
')}{\omega ^{\prime }\left( \omega ^{\prime }-\omega -i\eta \right) 
}\right].
\end{equation}%
%
From this, in the $\eta\to0$ limit we find
\begin{equation}
\text{Re}~\sigma _{xx}(\omega )=\frac{\chi
_{\mathrm{j}_x\mathrm{j}_x}''(\omega )}{\omega }. \label{CondNew}
\end{equation}%
%
We use the above formula in the main text to obtain the conductivity. 
It is gauge invariant at $q_x=0$ where the scalar potential does not contribute. 
Total charge is conserved at $q_x=0$ since it commutes with the Hamiltonian.

Local charge conservation, on the other hand, implies the $f$-sum rule  
\begin{figure}[h]
    \centering
    \includegraphics[width=0.85\linewidth]{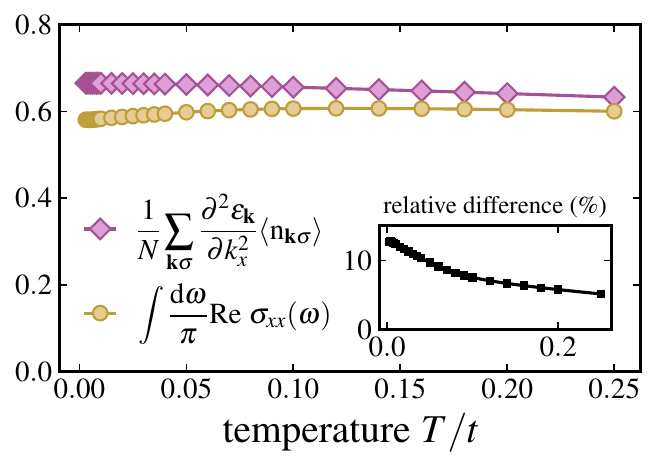}
    \caption{
    Comparison between the left-hand side (yellow circles) and right-hand side (pink diamonds) of \equref{eq:sum_rule_chijj}, as a function of temperature, at the QCP ($n=1.1827$) for $t'=t''=0$, $U=6t$.
    The inset shows the absolute value of the relative difference between the two.
    }
    \label{fig:sum_rule}
\end{figure}
%
\begin{align}
    \lim_{\omega\rightarrow 0}\lim_{q_x\rightarrow 0}\chi_{\mathrm{j}_x\mathrm{j}_x}(q_x,\omega) 
    &= \chi_{\mathrm{j}_x\mathrm{j}_x}(iq_n=0) \\ 
    &= \int \frac{%
d\omega ^{\prime }}{\pi }\frac{\chi''_{\mathrm{j}_x\mathrm{j}_x}(\omega ^{\prime })}{\omega ^{\prime }} \\
    &= \frac{1}{N}\sum_{\mathbf{k}\sigma}
    \frac{\partial^2 \epsilon_\mathbf{k}}{\partial k_x^2}
    \langle\mathrm{n}_{\mathbf{k}\sigma}\rangle.
    \label{eq:sum_rule_chijj}
\end{align}
%
%
While the high-frequency expansion of the charge susceptibility in TPSC+ leads to a divergent expression for the $f$-sum rule, the right-hand side of~\equref{eq:sum_rule_chijj} computed from the single-particle Green's function is close to the left-hand side.
\figref{fig:sum_rule} illustrates this.
Although \equref{eq:sum_rule_chijj} usually appears in the expression for the conductivity, our choice is instead to use the zero wave vector, zero-frequency current-current correlation function to ensure gauge invariance (\equref{eq:response_to_A}) and total charge conservation. 

While the relation between the high-frequency expansion of the charge susceptibility and the current correlation function is lost in TPSC+~\cite{gauvin-ndiaye_improved_2023}, using \equref{CondNew} to calculate $\sigma_{xx}$ ensures that the relation $\int\frac{\mathrm{d}\omega}{\omega}\mathrm{Re}~\sigma_{xx}(\omega)=\chi_{\mathrm{j}_x\mathrm{j}_x}(iq_n=0)$ is satisfied exactly.
\paragraph{Appendix: Kohn anomalies---}
%
\begin{figure}[h]
    \centering
    \includegraphics[width=0.8\linewidth]{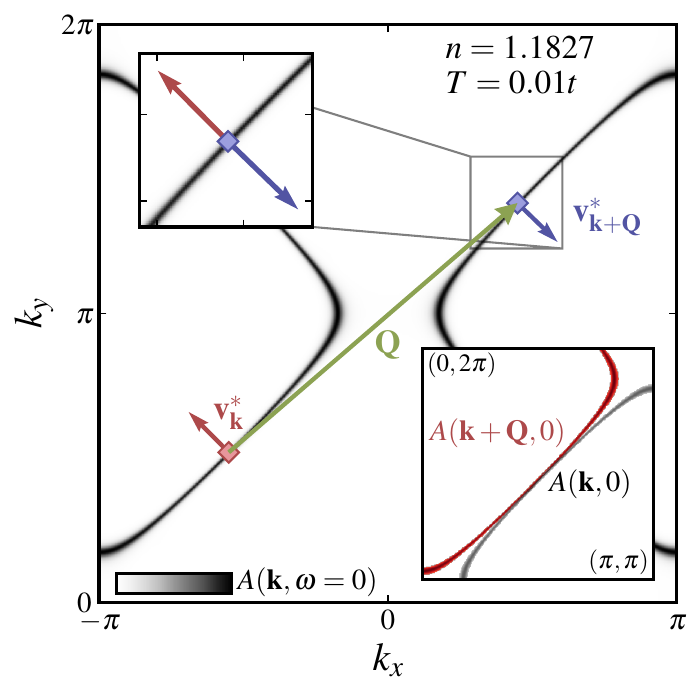}
    \caption{
    Zero-frequency spectral weight at the QCP ($n=1.1827$) for $t'=t''=0$, $U=6t$ and $T=0.01t$.
    The blue and red diamonds mark 2 points on the Fermi surface related by the incommensurate SDW wavevector $\mathbf{Q}_i=(\pi, \pi-\epsilon)$ (in green).
    The vectors at both points are the renormalized velocities evaluated at their respective points.
    The inset in the top left corner shows both velocities juxtaposed at the $\mathbf{k}+\mathbf{Q}_i$ point.
    The inset in the lower right corner shows the Fermi surface (black) and a Fermi surface translated by $\mathbf{Q}_i$ (red).
    }
    \label{fig:kohn}
\end{figure}
%
We are interested in the potential presence of Kohn anomalies associated with the Fermi surface \cite{Kohn_1959} to explain the peculiar temperature dependence of microscopic parameters.
We find these anomalies when two points on the Fermi surface separated by the wavevector $\mathbf{Q}$ have anti-parallel velocities \cite{Schafer_Katanin_Held_Toschi_2017}. 
To check whether Kohn anomalies are present, we check the cosine of the angle $\theta$ separating the velocities $\mathbf{v_k}$ and $\mathbf{v}_{\mathbf{k}+\mathbf{Q}_i}$ of the Fermi surface. 
Taking the bare dispersion relation ($\mathbf{v}_\mathbf{k}\propto \nabla_\mathbf{k}\epsilon_\mathbf{k}$), we find $\cos{\theta}\approx -0.99$.
The velocities are then almost anti-collinear.
This is strengthened if we instead use the renormalized dispersion ($\mathbf{v}^*_\mathbf{k}\propto \nabla_\mathbf{k}\epsilon_\mathbf{k}+\nabla_\mathbf{k}\mathrm{Re}\Sigma(\mathbf{k},\omega=0)$) where in this case $\cos\theta\approx -0.999$.
These velocities are illustrated in \figref{fig:kohn}.
We see them side by side in the inset in the top-left corner.
The same figure shows the almost perfect nesting using a Fermi surface displaced by $\mathbf{Q}$. The SM~\cite{SM} discusses the curvature of the Fermi surface. 

\end{document}

%% file: preamble.tex
\usepackage{graphicx}
\usepackage{grffile}
\usepackage{amsmath}
\usepackage{amssymb}
\usepackage{bm}
\usepackage{color}
\usepackage[dvipsnames]{xcolor}
\usepackage{amsmath,amssymb,amsfonts}
\usepackage{epsfig}
\usepackage{times}
\usepackage[colorlinks,bookmarks=false,citecolor=blue,linkcolor=blue,urlcolor=blue]{hyperref}
\usepackage{gensymb}
\usepackage[toc,page]{appendix}
\usepackage{float}

\usepackage{graphicx}
\graphicspath{ {./figures/} }
\newcommand{\figref}[1]{Fig.~\ref{#1}}
\newcommand{\equref}[1]{Eq.~(\ref{#1})}

%% file: bibli.bib
@article{gauvin-ndiaye_improved_2023,
	title = {Improved two-particle self-consistent approach for the single-band {Hubbard} model in two dimensions},
	volume = {108},
	url = {https://link.aps.org/doi/10.1103/PhysRevB.108.075144},
	doi = {10.1103/PhysRevB.108.075144},
	number = {7},
	urldate = {2023-08-18},
	journal = {Physical Review B},
	author = {Gauvin-Ndiaye, C. and Lahaie, C. and Vilk, Y. M. and Tremblay, A.-M. S.},
	month = aug,
	year = {2023},
	note = {Publisher: American Physical Society},
	pages = {075144},
}

@article{vilk_temblay_1997,
  title = {Non-{{Perturbative Many-Body Approach}} to the {{Hubbard Model}} and {{Single-Particle Pseudogap}}},
  author = {Vilk, Y. M. and Tremblay, A.-M. S.},
  year = {1997},
  month = nov,
  journal = {Journal de Physique I},
  volume = {7},
  number = {11},
  pages = {1309--1368},
  issn = {1155-4304, 1286-4862},
  doi = {10.1051/jp1:1997135},
  urldate = {2024-07-04},
}

@article{bergeron_optical_2011,
	title = {Optical and dc conductivity of the two-dimensional {Hubbard} model in the pseudogap regime and across the antiferromagnetic quantum critical point including vertex corrections},
	volume = {84},
	issn = {1098-0121, 1550-235X},
	url = {https://link.aps.org/doi/10.1103/PhysRevB.84.085128},
	doi = {10.1103/PhysRevB.84.085128},
	number = {8},
	urldate = {2023-01-28},
	journal = {Physical Review B},
	author = {Bergeron, Dominic and Hankevych, Vasyl and Kyung, Bumsoo and Tremblay, A.-M. S.},
	month = aug,
	year = {2011},
	pages = {085128},
}

@article{gauvin_2022_resilient_fermi_liquid,
  title = {Resilient Fermi Liquid and Strength of Correlations near an Antiferromagnetic Quantum Critical Point},
  author = {Gauvin-Ndiaye, C. and Setrakian, M. and Tremblay, A.-M. S.},
  journal = {Phys. Rev. Lett.},
  volume = {128},
  issue = {8},
  pages = {087001},
  numpages = {6},
  year = {2022},
  month = {Feb},
  publisher = {American Physical Society},
  doi = {10.1103/PhysRevLett.128.087001},
  url = {https://link.aps.org/doi/10.1103/PhysRevLett.128.087001}
}

@article{bergeron_algorithms_2016,
	title = {Algorithms for optimized maximum entropy and diagnostic tools for analytic continuation},
	volume = {94},
	issn = {2470-0045, 2470-0053},
	url = {https://link.aps.org/doi/10.1103/PhysRevE.94.023303},
	doi = {10.1103/PhysRevE.94.023303},
	number = {2},
	urldate = {2022-04-04},
	journal = {Physical Review E},
	author = {Bergeron, Dominic and Tremblay, A.-M. S.},
	month = aug,
	year = {2016},
	pages = {023303},
}

@article{paul2003,
  title = {Thermal Transport for Many-Body Tight-Binding Models},
  author = {Paul, Indranil and Kotliar, Gabriel},
  year = {2003},
  month = mar,
  journal = {Physical Review B},
  volume = {67},
  number = {11},
  pages = {115131},
  publisher = {American Physical Society},
  doi = {10.1103/PhysRevB.67.115131},
}

@article{arsenault_transport_2013,
	title = {Transport functions for hypercubic and {Bethe} lattices},
	volume = {88},
	issn = {1098-0121, 1550-235X},
	url = {https://link.aps.org/doi/10.1103/PhysRevB.88.205109},
	doi = {10.1103/PhysRevB.88.205109},
	number = {20},
	urldate = {2021-12-14},
	journal = {Physical Review B},
	author = {Arsenault, Louis-François and Tremblay, A.-M. S.},
	month = nov,
	year = {2013},
	pages = {205109},
}

@article{reymbaut_maximum_2017,
	title = {Maximum entropy analytic continuation for frequency-dependent transport coefficients with nonpositive spectral weight},
	volume = {95},
	issn = {2469-9950, 2469-9969},
	url = {https://link.aps.org/doi/10.1103/PhysRevB.95.121104},
	doi = {10.1103/PhysRevB.95.121104},
	number = {12},
	urldate = {2022-04-04},
	journal = {Physical Review B},
	author = {Reymbaut, A. and Gagnon, A.-M. and Bergeron, D. and Tremblay, A.-M. S.},
	month = mar,
	year = {2017},
	pages = {121104},
}

@article{Dare1996Crossover,
  title = {Crossover from two- to three-dimensional critical behavior for nearly antiferromagnetic itinerant electrons},
  author = {Dar\'e, Anne-Marie and Vilk, Y. M. and Tremblay, A. -M. S.},
  journal = {Phys. Rev. B},
  volume = {53},
  issue = {21},
  pages = {14236--14251},
  numpages = {0},
  year = {1996},
  month = {Jun},
  publisher = {American Physical Society},
  doi = {10.1103/PhysRevB.53.14236},
  url = {https://link.aps.org/doi/10.1103/PhysRevB.53.14236}
}

@Article{vilk2024antiferromagnetic,
  author    = {Vilk, YM and Lahaie, Camille and Tremblay, A-MS},
  journal   = {Physical Review B},
  title     = {Antiferromagnetic pseudogap in the two-dimensional Hubbard model deep in the renormalized classical regime},
  year      = {2024},
  number    = {12},
  pages     = {125154},
  volume    = {110},
  publisher = {APS},
  url       = {https://journals.aps.org/prb/abstract/10.1103/PhysRevB.110.125154},
}

@article{pakhira2015,
  title = {Absence of a quantum limit to charge diffusion in bad metals},
  author = {Pakhira, Nandan and McKenzie, Ross H.},
  journal = {Phys. Rev. B},
  volume = {91},
  issue = {7},
  pages = {075124},
  numpages = {10},
  year = {2015},
  month = {Feb},
  publisher = {American Physical Society},
  doi = {10.1103/PhysRevB.91.075124},
  url = {https://link.aps.org/doi/10.1103/PhysRevB.91.075124}
}

@article{gurvitch_1987,
 title = {Resistivity of {La}$_{1.825}${Sr}$_{0.175}${CuO}$_{4}$ and {YBa}$_{2}${Cu}$_{3}${O}$_{7}$ to 1100 K: Absence of saturation and its implications},
  author = {Gurvitch, M. and Fiory, A. T.},
  journal = {Phys. Rev. Lett.},
  volume = {59},
  issue = {12},
  pages = {1337--1340},
  numpages = {0},
  year = {1987},
  month = {Sep},
  publisher = {American Physical Society},
  doi = {10.1103/PhysRevLett.59.1337},
  url = {https://link.aps.org/doi/10.1103/PhysRevLett.59.1337}
}

@Article{Brown_Mitra_Bakr_2018,
  author       = {Brown, Peter T. and Mitra, Debayan and Guardado-Sanchez, Elmer and Nourafkan, Reza and Reymbaut, Alexis and Hébert, Charles-David and Bergeron, Simon and Tremblay, A.-M. S. and Kokalj, Jure and Huse, David A. and et al.},
  journal      = {Science},
  title        = {Bad metallic transport in a cold atom {F}ermi-{H}ubbard system},
  year         = {2018},
  issn         = {0036-8075},
  doi          = {10.1126/science.aat4134},
  url          = {http://science.sciencemag.org/content/early/2018/12/06/science.aat4134},
}

@Article{Simkovic_Rossi_Georges_Ferrero_2024,
  author       = {Šimkovic, Fedor and Rossi, Riccardo and Georges, Antoine and Ferrero, Michel},
  journal      = {Science},
  title        = {Origin and fate of the pseudogap in the doped Hubbard model},
  year         = {2024},
  month        = sept,
  number       = {6715},
  pages        = {eade9194},
  volume       = {385},
    doi          = {10.1126/science.ade9194},
  publisher    = {American Association for the Advancement of Science},
}

@article{Vilk_Tremblay_1995, title={Destruction of the fermi liquid by spin fluctuations in two dimensions}, volume={56}, ISSN={0022-3697}, DOI={https://doi.org/10.1016/0022-3697(95)00168-9},  number={12}, journal={Journal of Physics and Chemistry of Solids}, author={Vilk, Y. M. and Tremblay, A.-M. S.}, year={1995}, pages={1769–1771}
 }

@article{Kyung2004, 
  title = {Pseudogap and Spin Fluctuations in the Normal State of the Electron-Doped Cuprates},
  author = {Kyung, B. and Hankevych, V. and Dar\'e, A.-M. and Tremblay, A.-M. S.},
  journal = {Phys. Rev. Lett.},
  volume = {93},
  issue = {14},
  pages = {147004},
  numpages = {4},
  year = {2004},
  month = {Sep},
  publisher = {American Physical Society},
  doi = {10.1103/PhysRevLett.93.147004},
  url = {https://link.aps.org/doi/10.1103/PhysRevLett.93.147004}
}

@article{senechal_2004_pseudogap,
  title = {Hot Spots and Pseudogaps for Hole- and Electron-Doped High-Temperature Superconductors},
  author = {S\'en\'echal, David and Tremblay, A.-M. S.},
  journal = {Phys. Rev. Lett.},
  volume = {92},
  issue = {12},
  pages = {126401},
  numpages = {4},
  year = {2004},
  month = {Mar},
  publisher = {American Physical Society},
  doi = {10.1103/PhysRevLett.92.126401},
  url = {https://link.aps.org/doi/10.1103/PhysRevLett.92.126401}
}

@article{kyung2006,
  title = {Pseudogap induced by short-range spin correlations in a doped Mott insulator},
  author = {Kyung, B. and Kancharla, S. S. and S\'en\'echal, D. and Tremblay, A.-M. S. and Civelli, M. and Kotliar, G.},
  journal = {Phys. Rev. B},
  volume = {73},
  issue = {16},
  pages = {165114},
  numpages = {6},
  year = {2006},
  month = {Apr},
  publisher = {American Physical Society},
  doi = {10.1103/PhysRevB.73.165114},
  url = {https://link.aps.org/doi/10.1103/PhysRevB.73.165114}
}

@article{roy2008scaling,
  title={Scaling and commensurate-incommensurate crossover for the d= 2, z= 2 quantum critical point of itinerant antiferromagnets},
  author={Roy, S{\'e}bastien and Tremblay, A-MS},
  journal={EPL (Europhysics Letters)},
  volume={84},
  number={3},
  pages={37013},
  year={2008},
  url = {https://iopscience.iop.org/article/10.1209/0295-5075/84/37013/meta}
}

@article{moriya1990antiferromagnetic,
  title={Antiferromagnetic spin fluctuations and superconductivity in two-dimensional metals—a possible model for high Tc oxides},
  author={Moriya, T{\^o}ru and Takahashi, Yoshinori and Ueda, Kazuo},
  journal={Journal of the Physical Society of Japan},
  volume={59},
  number={8},
  pages={2905--2915},
  year={1990},
  publisher={The Physical Society of Japan},
  url={https://www.jstage.jst.go.jp/article/jpsj1946/59/8/59_8_2905/_article/-char/ja/}
}

@article{millis1993effect,
  title={Effect of a nonzero temperature on quantum critical points in itinerant fermion systems},
  author={Millis, AJ},
  journal={Physical Review B},
  volume={48},
  number={10},
  pages={7183},
  year={1993},
  publisher={APS},
  url={https://journals.aps.org/prb/pdf/10.1103/PhysRevB.48.7183}
}

@article{Bergeron2012QCP,
  title = {Breakdown of Fermi liquid behavior at the $(\ensuremath{\pi},\ensuremath{\pi})=2{k}_{F}$ spin-density wave quantum-critical point: The case of electron-doped cuprates},
  author = {Bergeron, Dominic and Chowdhury, Debanjan and Punk, Matthias and Sachdev, Subir and Tremblay, A.-M. S.},
  journal = {Phys. Rev. B},
  volume = {86},
  issue = {15},
  pages = {155123},
  numpages = {17},
  year = {2012},
  month = {Oct},
  publisher = {American Physical Society},
  doi = {10.1103/PhysRevB.86.155123},
  url = {https://link.aps.org/doi/10.1103/PhysRevB.86.155123}
}

@article{Kendrick_Kale_Gang_Deters_Lebrat_Young_Greiner_2025, title={Pseudogap in a Fermi-Hubbard quantum simulator},
 url={http://arxiv.org/abs/2509.18075},
 DOI={10.48550/arXiv.2509.18075}, 
 journal={},
 note={arXiv:2509.18075 [cond-mat]},
 number={arXiv:2509.18075},
 publisher={arXiv},
 author={Kendrick, Lev Haldar and Kale, Anant and Gang, Youqi and Deters, Alexander Dennisovich and Lebrat, Martin and Young, Aaron W. and Greiner, Markus},
 year={2025},
 month={sept}, }

@article{brown_2019,
author = {Peter T. Brown  and Debayan Mitra  and Elmer Guardado-Sanchez  and Reza Nourafkan  and Alexis Reymbaut  and Charles-David Hébert  and Simon Bergeron  and A.-M. S. Tremblay  and Jure Kokalj  and David A. Huse  and Peter Schauß  and Waseem S. Bakr },
title = {Bad metallic transport in a cold atom Fermi-Hubbard system},
journal = {Science},
volume = {363},
number = {6425},
pages = {379-382},
year = {2019},
doi = {10.1126/science.aat4134},
URL = {https://www.science.org/doi/abs/10.1126/science.aat4134},
eprint = {https://www.science.org/doi/pdf/10.1126/science.aat4134},
}

@article{brown2020,
  title = {Angle-Resolved Photoemission Spectroscopy of a {{Fermi}}--{{Hubbard}} System},
  author = {Brown, Peter T. and {Guardado-Sanchez}, Elmer and Spar, Benjamin M. and Huang, Edwin W. and Devereaux, Thomas P. and Bakr, Waseem S.},
  year = 2020,
  month = jan,
  journal = {Nature Physics},
  volume = {16},
  number = {1},
  pages = {26--31},
  issn = {1745-2481},
  doi = {10.1038/s41567-019-0696-0},
}

@article{anderson2019,
  title = {Conductivity Spectrum of Ultracold Atoms in an Optical Lattice},
  author = {Anderson, Rhys and Wang, Fudong and Xu, Peihang and Venu, Vijin and Trotzky, Stefan and Chevy, Fr\'ed\'eric and Thywissen, Joseph H.},
  journal = {Phys. Rev. Lett.},
  volume = {122},
  issue = {15},
  pages = {153602},
  numpages = {7},
  year = {2019},
  month = {Apr},
  publisher = {American Physical Society},
  doi = {10.1103/PhysRevLett.122.153602},
  url = {https://link.aps.org/doi/10.1103/PhysRevLett.122.153602}
}

@article{
chalopin2025,
author = {Thomas Chalopin  and Petar Bojović  and Si Wang  and Titus Franz  and Aritra Sinha  and Zhenjiu Wang  and Dominik Bourgund  and Johannes Obermeyer  and Fabian Grusdt  and Annabelle Bohrdt  and Lode Pollet  and Alexander Wietek  and Antoine Georges  and Timon Hilker  and Immanuel Bloch },
title = {Observation of emergent scaling of spin–charge correlations at the onset of the pseudogap},
journal = {Proceedings of the National Academy of Sciences},
volume = {123},
number = {4},
pages = {e2525539123},
year = {2026},
doi = {10.1073/pnas.2525539123},
URL = {https://www.pnas.org/doi/abs/10.1073/pnas.2525539123},
eprint = {https://www.pnas.org/doi/pdf/10.1073/pnas.2525539123},
}

@article{cocchi2016,
  title = {Equation of State of the Two-Dimensional Hubbard Model},
  author = {Cocchi, Eugenio and Miller, Luke A. and Drewes, Jan H. and Koschorreck, Marco and Pertot, Daniel and Brennecke, Ferdinand and K\"ohl, Michael},
  journal = {Phys. Rev. Lett.},
  volume = {116},
  issue = {17},
  pages = {175301},
  numpages = {5},
  year = {2016},
  month = {Apr},
  publisher = {American Physical Society},
  doi = {10.1103/PhysRevLett.116.175301},
  url = {https://link.aps.org/doi/10.1103/PhysRevLett.116.175301}
}

@article{pasqualetti2024,
  title = {Equation of State and Thermometry of the 2D $\mathrm{SU}(N)$ Fermi-Hubbard Model},
  author = {Pasqualetti, G. and Bettermann, O. and Darkwah Oppong, N. and Ibarra-Garc\'{\i}a-Padilla, E. and Dasgupta, S. and Scalettar, R. T. and Hazzard, K. R. A. and Bloch, I. and F\"olling, S.},
  journal = {Phys. Rev. Lett.},
  volume = {132},
  issue = {8},
  pages = {083401},
  numpages = {8},
  year = {2024},
  month = {Feb},
  publisher = {American Physical Society},
  doi = {10.1103/PhysRevLett.132.083401},
  url = {https://link.aps.org/doi/10.1103/PhysRevLett.132.083401}
}

@misc{richard2025,
     title={Observation of Magnon-Polarons in the Fermi-Hubbard Model}, 
     author={Max L. Prichard and Zengli Ba and Ivan Morera and Benjamin M. Spar and David A. Huse and Eugene Demler and Waseem S. Bakr},
    year={2025},
    eprint={2502.06757},
    archivePrefix={arXiv},
    primaryClass={cond-mat.quant-gas},
    url={https://arxiv.org/abs/2502.06757}, 
}

@article{doiron2009,
  title = {Correlation between linear resistivity and ${T}_{c}$ in the Bechgaard salts and the pnictide superconductor $\text{Ba}{({\text{Fe}}_{1\ensuremath{-}x}{\text{Co}}_{x})}_{2}{\text{As}}_{2}$},
  author = {Doiron-Leyraud, Nicolas and Auban-Senzier, Pascale and Ren\'e de Cotret, Samuel and Bourbonnais, Claude and J\'erome, Denis and Bechgaard, Klaus and Taillefer, Louis},
  journal = {Phys. Rev. B},
  volume = {80},
  issue = {21},
  pages = {214531},
  numpages = {5},
  year = {2009},
  month = {Dec},
  publisher = {American Physical Society},
  doi = {10.1103/PhysRevB.80.214531},
  url = {https://link.aps.org/doi/10.1103/PhysRevB.80.214531}
}

@article{doiron-leyraudLinearTScatteringPairing2010,
  title = {Linear-{{T}} Scattering and Pairing from Antiferromagnetic Fluctuations in the ({{TMTSF}}){{2X}} Organic Superconductors},
  author = {{Doiron-Leyraud}, N. and {Ren{\'e} de Cotret}, S. and Sedeki, A. and Bourbonnais, C. and Taillefer, L. and {Auban-Senzier}, P. and J{\'e}rome, D. and Bechgaard, K.},
  year = 2010,
  month = nov,
  journal = {The European Physical Journal B},
  volume = {78},
  number = {1},
  pages = {23--36},
  issn = {1434-6036},
  doi = {10.1140/epjb/e2010-10571-4},
}

@article{yeHoppingFrustrationinducedFlat2024,
  title = {Hopping Frustration-Induced Flat Band and Strange Metallicity in a Kagome Metal},
  author = {Ye, Linda and Fang, Shiang and Kang, Mingu and Kaufmann, Josef and Lee, Yonghun and John, Caolan and Neves, Paul M. and Zhao, S. Y. Frank and Denlinger, Jonathan and Jozwiak, Chris and Bostwick, Aaron and Rotenberg, Eli and Kaxiras, Efthimios and Bell, David C. and Janson, Oleg and Comin, Riccardo and Checkelsky, Joseph G.},
  year = 2024,
  month = apr,
  journal = {Nature Physics},
  volume = {20},
  number = {4},
  pages = {610--614},
  issn = {1745-2481},
  doi = {10.1038/s41567-023-02360-5},
}

@article{meierStrangeWayStrangeMetal2024,
  title = {A Strange Way to Get a Strange Metal},
  author = {Meier, William R.},
  year = 2024,
  month = apr,
  journal = {Nature Physics},
  volume = {20},
  number = {4},
  pages = {541--542},
  issn = {1745-2481},
  doi = {10.1038/s41567-024-02416-0}
}

@article{fang2009,
  title = {Roles of multiband effects and electron-hole asymmetry in the superconductivity and normal-state properties of $\text{Ba}{({\text{Fe}}_{1\ensuremath{-}x}{\text{Co}}_{x})}_{2}{\text{As}}_{2}$},
  author = {Fang, Lei and Luo, Huiqian and Cheng, Peng and Wang, Zhaosheng and Jia, Ying and Mu, Gang and Shen, Bing and Mazin, I. I. and Shan, Lei and Ren, Cong and Wen, Hai-Hu},
  journal = {Phys. Rev. B},
  volume = {80},
  issue = {14},
  pages = {140508},
  numpages = {4},
  year = {2009},
  month = {Oct},
  publisher = {American Physical Society},
  doi = {10.1103/PhysRevB.80.140508},
  url = {https://link.aps.org/doi/10.1103/PhysRevB.80.140508}
}

@article{yuanScalingStrangemetalScattering2022,
  title = {Scaling of the Strange-Metal Scattering in Unconventional Superconductors},
  author = {Yuan, Jie and Chen, Qihong and Jiang, Kun and Feng, Zhongpei and Lin, Zefeng and Yu, Heshan and He, Ge and Zhang, Jinsong and Jiang, Xingyu and Zhang, Xu and Shi, Yujun and Zhang, Yanmin and Qin, Mingyang and Cheng, Zhi Gang and Tamura, Nobumichi and Yang, Yi-feng and Xiang, Tao and Hu, Jiangping and Takeuchi, Ichiro and Jin, Kui and Zhao, Zhongxian},
  year = 2022,
  month = feb,
  journal = {Nature},
  volume = {602},
  number = {7897},
  pages = {431--436},
  issn = {1476-4687},
  doi = {10.1038/s41586-021-04305-5},
}

@article{jinLinkSpinFluctuations2011,
  title = {Link between Spin Fluctuations and Electron Pairing in Copper Oxide Superconductors},
  author = {Jin, K. and Butch, N. P. and Kirshenbaum, K. and Paglione, J. and Greene, R. L.},
  year = 2011,
  month = aug,
  journal = {Nature},
  volume = {476},
  number = {7358},
  pages = {73--75},
  issn = {1476-4687},
  doi = {10.1038/nature10308},
  }

@article{legrosUniversalTlinearResistivity2019,
  title = {Universal {{T-linear}} Resistivity and {{Planckian}} Dissipation in Overdoped Cuprates},
  author = {Legros, A. and Benhabib, S. and Tabis, W. and Lalibert{\'e}, F. and Dion, M. and Lizaire, M. and Vignolle, B. and Vignolles, D. and Raffy, H. and Li, Z. Z. and {Auban-Senzier}, P. and {Doiron-Leyraud}, N. and Fournier, P. and Colson, D. and Taillefer, L. and Proust, C.},
  year = 2019,
  month = feb,
  journal = {Nature Physics},
  volume = {15},
  number = {2},
  pages = {142--147},
  issn = {1745-2481},
  doi = {10.1038/s41567-018-0334-2},
}

@article{cooper2009,
author = {R. A. Cooper  and Y. Wang  and B. Vignolle  and O. J. Lipscombe  and S. M. Hayden  and Y. Tanabe  and T. Adachi  and Y. Koike  and M. Nohara  and H. Takagi  and Cyril Proust  and N. E. Hussey },
title = {Anomalous Criticality in the Electrical Resistivity of $\mathrm{La}_{2-x}\mathrm{Sr}_x\mathrm{CuO}_4$},
journal = {Science},
volume = {323},
number = {5914},
pages = {603-607},
year = {2009},
doi = {10.1126/science.1165015},
URL = {https://www.science.org/doi/abs/10.1126/science.1165015},
eprint = {https://www.science.org/doi/pdf/10.1126/science.1165015},
}

@article{takagi1992,
  title = {{Systematic evolution of temperature-dependent resistivity in ${\mathrm{La}}_{2\mathrm{\ensuremath{-}}\mathit{x}}$${\mathrm{Sr}}_{\mathit{x}}$${\mathrm{CuO}}_{4}$}},
  author = {Takagi, H. and Batlogg, B. and Kao, H. L. and Kwo, J. and Cava, R. J. and Krajewski, J. J. and Peck, W. F.},
  journal = {Phys. Rev. Lett.},
  volume = {69},
  issue = {20},
  pages = {2975--2978},
  numpages = {0},
  year = {1992},
  month = {Nov},
  publisher = {American Physical Society},
  doi = {10.1103/PhysRevLett.69.2975},
  url = {https://link.aps.org/doi/10.1103/PhysRevLett.69.2975}
}

@article{martin1990,
  title = {Normal-state transport properties of ${\mathrm{Bi}}_{2+\mathit{x}}$${\mathrm{Sr}}_{2\mathrm{\ensuremath{-}}\mathit{y}}$${\mathrm{CuO}}_{6+\mathrm{\ensuremath{\delta}}}$ crystals},
  author = {Martin, S. and Fiory, A. T. and Fleming, R. M. and Schneemeyer, L. F. and Waszczak, J. V.},
  journal = {Phys. Rev. B},
  volume = {41},
  issue = {1},
  pages = {846--849},
  numpages = {0},
  year = {1990},
  month = {Jan},
  publisher = {American Physical Society},
  doi = {10.1103/PhysRevB.41.846},
  url = {https://link.aps.org/doi/10.1103/PhysRevB.41.846}
}

@article{mackenzie1996,
  title = {Normal-state magnetotransport in superconducting ${\mathrm{Tl}}_{2}$${\mathrm{Ba}}_{2}$${\mathrm{CuO}}_{6+\mathrm{\ensuremath{\delta}}}$ to millikelvin temperatures},
  author = {Mackenzie, A. P. and Julian, S. R. and Sinclair, D. C. and Lin, C. T.},
  journal = {Phys. Rev. B},
  volume = {53},
  issue = {9},
  pages = {5848--5855},
  numpages = {0},
  year = {1996},
  month = {Mar},
  publisher = {American Physical Society},
  doi = {10.1103/PhysRevB.53.5848},
  url = {https://link.aps.org/doi/10.1103/PhysRevB.53.5848}
}

@article{fournierInsulatorMetal1998,
  title = {{Insulator-Metal Crossover near Optimal Doping in ${\mathrm{Pr}}_{2\ensuremath{-}\mathit{x}}{\mathrm{Ce}}_{\mathit{x}}{\mathrm{CuO}}_{4}$: Anomalous Normal-State Low Temperature Resistivity}},
  author = {Fournier, P. and Mohanty, P. and Maiser, E. and Darzens, S. and Venkatesan, T. and Lobb, C. J. and Czjzek, G. and Webb, R. A. and Greene, R. L.},
  journal = {Phys. Rev. Lett.},
  volume = {81},
  issue = {21},
  pages = {4720--4723},
  numpages = {0},
  year = {1998},
  month = {Nov},
  publisher = {American Physical Society},
  doi = {10.1103/PhysRevLett.81.4720},
  url = {https://link.aps.org/doi/10.1103/PhysRevLett.81.4720}
}

@article{custersBreakupHeavyElectrons2003,
  title = {The Break-up of Heavy Electrons at a Quantum Critical Point},
  author = {Custers, J. and Gegenwart, P. and Wilhelm, H. and Neumaier, K. and Tokiwa, Y. and Trovarelli, O. and Geibel, C. and Steglich, F. and P{\'e}pin, C. and Coleman, P.},
  year = 2003,
  month = jul,
  journal = {Nature},
  volume = {424},
  number = {6948},
  pages = {524--527},
  issn = {1476-4687},
  doi = {10.1038/nature01774},
}

@article{
martelli2019,
author = {Valentina Martelli  and Ang Cai  and Emilian M. Nica  and Mathieu Taupin  and Andrey Prokofiev  and Chia-Chuan Liu  and Hsin-Hua Lai  and Rong Yu  and Kevin Ingersent  and Robert Küchler  and André M. Strydom  and Diana Geiger  and Jonathan Haenel  and Julio Larrea  and Qimiao Si  and Silke Paschen },
title = {Sequential localization of a complex electron fluid},
journal = {Proceedings of the National Academy of Sciences},
volume = {116},
number = {36},
pages = {17701-17706},
year = {2019},
doi = {10.1073/pnas.1908101116},
URL = {https://www.pnas.org/doi/abs/10.1073/pnas.1908101116},
eprint = {https://www.pnas.org/doi/pdf/10.1073/pnas.1908101116},
}

@Article{taupin2022,
AUTHOR = {Taupin, Mathieu and Paschen, Silke},
TITLE = {Are Heavy Fermion Strange Metals Planckian?},
JOURNAL = {Crystals},
VOLUME = {12},
YEAR = {2022},
NUMBER = {2},
ARTICLE-NUMBER = {251},
URL = {https://www.mdpi.com/2073-4352/12/2/251},
ISSN = {2073-4352},
DOI = {10.3390/cryst12020251}
}

@article{hlubinaresistivity1995,
  title = {Resistivity as a function of temperature for models with hot spots on the Fermi surface},
  author = {Hlubina, R. and Rice, T. M.},
  journal = {Phys. Rev. B},
  volume = {51},
  issue = {14},
  pages = {9253--9260},
  numpages = {0},
  year = {1995},
  month = {Apr},
  publisher = {American Physical Society},
  doi = {10.1103/PhysRevB.51.9253},
  url = {https://link.aps.org/doi/10.1103/PhysRevB.51.9253}
}

@article{chowdhurySYK2022,
  title = {Sachdev-Ye-Kitaev models and beyond: Window into non-Fermi liquids},
  author = {Chowdhury, Debanjan and Georges, Antoine and Parcollet, Olivier and Sachdev, Subir},
  journal = {Rev. Mod. Phys.},
  volume = {94},
  issue = {3},
  pages = {035004},
  numpages = {78},
  year = {2022},
  month = {Sep},
  publisher = {American Physical Society},
  doi = {10.1103/RevModPhys.94.035004},
  url = {https://link.aps.org/doi/10.1103/RevModPhys.94.035004}
}

@article{ferreroPseudogap2009,
  title = {Pseudogap opening and formation of Fermi arcs as an orbital-selective Mott transition in momentum space},
  author = {Ferrero, Michel and Cornaglia, Pablo S. and De Leo, Lorenzo and Parcollet, Olivier and Kotliar, Gabriel and Georges, Antoine},
  journal = {Phys. Rev. B},
  volume = {80},
  issue = {6},
  pages = {064501},
  numpages = {21},
  year = {2009},
  month = {Aug},
  publisher = {American Physical Society},
  doi = {10.1103/PhysRevB.80.064501},
  url = {https://link.aps.org/doi/10.1103/PhysRevB.80.064501}
}

@article{gullMillis2013,
  title = {Superconducting and pseudogap effects on the interplane conductivity and Raman scattering cross section in the two-dimensional Hubbard model},
  author = {Gull, E. and Millis, A. J.},
  journal = {Phys. Rev. B},
  volume = {88},
  issue = {7},
  pages = {075127},
  numpages = {12},
  year = {2013},
  month = {Aug},
  publisher = {American Physical Society},
  doi = {10.1103/PhysRevB.88.075127},
  url = {https://link.aps.org/doi/10.1103/PhysRevB.88.075127}
}

@article{fratinoPseudogap2016,
  title = {Pseudogap and superconductivity in two-dimensional doped charge-transfer insulators},
  author = {Fratino, L. and S\'emon, P. and Sordi, G. and Tremblay, A.-M. S.},
  journal = {Phys. Rev. B},
  volume = {93},
  issue = {24},
  pages = {245147},
  numpages = {6},
  year = {2016},
  month = {Jun},
  publisher = {American Physical Society},
  doi = {10.1103/PhysRevB.93.245147},
  url = {https://link.aps.org/doi/10.1103/PhysRevB.93.245147}
}

@article{huscroft2001,
  title = {Pseudogaps in the 2D Hubbard Model},
  author = {Huscroft, C. and Jarrell, M. and Maier, Th. and Moukouri, S. and Tahvildarzadeh, A. N.},
  journal = {Phys. Rev. Lett.},
  volume = {86},
  issue = {1},
  pages = {139--142},
  numpages = {0},
  year = {2001},
  month = {Jan},
  publisher = {American Physical Society},
  doi = {10.1103/PhysRevLett.86.139},
  url = {https://link.aps.org/doi/10.1103/PhysRevLett.86.139}
}

@article{macridinpseudogap2006,
  title = {Pseudogap and Antiferromagnetic Correlations in the Hubbard Model},
  author = {Macridin, Alexandru and Jarrell, M. and Maier, Thomas and Kent, P. R. C. and D'Azevedo, Eduardo},
  journal = {Phys. Rev. Lett.},
  volume = {97},
  issue = {3},
  pages = {036401},
  numpages = {4},
  year = {2006},
  month = {Jul},
  publisher = {American Physical Society},
  doi = {10.1103/PhysRevLett.97.036401},
  url = {https://link.aps.org/doi/10.1103/PhysRevLett.97.036401}
}

@article{stanescuFermiArcs2006,
  title = {Fermi arcs and hidden zeros of the Green function in the pseudogap state},
  author = {Stanescu, Tudor D. and Kotliar, Gabriel},
  journal = {Phys. Rev. B},
  volume = {74},
  issue = {12},
  pages = {125110},
  numpages = {6},
  year = {2006},
  month = {Sep},
  publisher = {American Physical Society},
  doi = {10.1103/PhysRevB.74.125110},
  url = {https://link.aps.org/doi/10.1103/PhysRevB.74.125110}
}

@article{sakaiEvolution2009,
  title = {Evolution of Electronic Structure of Doped Mott Insulators: Reconstruction of Poles and Zeros of Green's Function},
  author = {Sakai, Shiro and Motome, Yukitoshi and Imada, Masatoshi},
  journal = {Phys. Rev. Lett.},
  volume = {102},
  issue = {5},
  pages = {056404},
  numpages = {4},
  year = {2009},
  month = {Feb},
  publisher = {American Physical Society},
  doi = {10.1103/PhysRevLett.102.056404},
  url = {https://link.aps.org/doi/10.1103/PhysRevLett.102.056404}
}

@article{sordiStrongCoupling2012,
  title = {Strong Coupling Superconductivity, Pseudogap, and Mott Transition},
  author = {Sordi, G. and S\'emon, P. and Haule, K. and Tremblay, A.-M. S.},
  journal = {Phys. Rev. Lett.},
  volume = {108},
  issue = {21},
  pages = {216401},
  numpages = {6},
  year = {2012},
  month = {May},
  publisher = {American Physical Society},
  doi = {10.1103/PhysRevLett.108.216401},
  url = {https://link.aps.org/doi/10.1103/PhysRevLett.108.216401}
}

@article{sordiCAxis2013,
  title = {$c$-axis resistivity, pseudogap, superconductivity, and Widom line in doped Mott insulators},
  author = {Sordi, G. and S\'emon, P. and Haule, K. and Tremblay, A.-M. S.},
  journal = {Phys. Rev. B},
  volume = {87},
  issue = {4},
  pages = {041101},
  numpages = {5},
  year = {2013},
  month = {Jan},
  publisher = {American Physical Society},
  doi = {10.1103/PhysRevB.87.041101},
  url = {https://link.aps.org/doi/10.1103/PhysRevB.87.041101}
}

@article{gullMillisLetters2013,
  title = {Superconductivity and the Pseudogap in the Two-Dimensional Hubbard Model},
  author = {Gull, Emanuel and Parcollet, Olivier and Millis, Andrew J.},
  journal = {Phys. Rev. Lett.},
  volume = {110},
  issue = {21},
  pages = {216405},
  numpages = {5},
  year = {2013},
  month = {May},
  publisher = {American Physical Society},
  doi = {10.1103/PhysRevLett.110.216405},
  url = {https://link.aps.org/doi/10.1103/PhysRevLett.110.216405}
}

@article{reymbautPseudogap2019,
  title = {Pseudogap, van Hove singularity, maximum in entropy, and specific heat for hole-doped Mott insulators},
  author = {Reymbaut, A. and Bergeron, S. and Garioud, R. and Th\'enault, M. and Charlebois, M. and S\'emon, P. and Tremblay, A.-M. S.},
  journal = {Phys. Rev. Res.},
  volume = {1},
  issue = {2},
  pages = {023015},
  numpages = {6},
  year = {2019},
  month = {Sep},
  publisher = {American Physical Society},
  doi = {10.1103/PhysRevResearch.1.023015},
  url = {https://link.aps.org/doi/10.1103/PhysRevResearch.1.023015}
}

@article{dahsPseudogap2019,
  title = {Pseudogap transition within the superconducting phase in the three-band Hubbard model},
  author = {Dash, S. S. and S\'en\'echal, D.},
  journal = {Phys. Rev. B},
  volume = {100},
  issue = {21},
  pages = {214509},
  numpages = {9},
  year = {2019},
  month = {Dec},
  publisher = {American Physical Society},
  doi = {10.1103/PhysRevB.100.214509},
  url = {https://link.aps.org/doi/10.1103/PhysRevB.100.214509}
}

@article{walshPrediction2022,
  title = {Prediction of anomalies in the velocity of sound for the pseudogap of hole-doped cuprates},
  author = {Walsh, C. and Charlebois, M. and S\'emon, P. and Sordi, G. and Tremblay, A.-M. S.},
  journal = {Phys. Rev. B},
  volume = {106},
  issue = {23},
  pages = {235134},
  numpages = {6},
  year = {2022},
  month = {Dec},
  publisher = {American Physical Society},
  doi = {10.1103/PhysRevB.106.235134},
  url = {https://link.aps.org/doi/10.1103/PhysRevB.106.235134}
}

@article{sordiSpecificHeat2019,
  title = {Specific heat maximum as a signature of Mott physics in the two-dimensional Hubbard model},
  author = {Sordi, G. and Walsh, C. and S\'emon, P. and Tremblay, A.-M. S.},
  journal = {Phys. Rev. B},
  volume = {100},
  issue = {12},
  pages = {121105},
  numpages = {6},
  year = {2019},
  month = {Sep},
  publisher = {American Physical Society},
  doi = {10.1103/PhysRevB.100.121105},
  url = {https://link.aps.org/doi/10.1103/PhysRevB.100.121105}
}

@article{walshCritical2019,
  title = {Critical opalescence across the doping-driven Mott transition in optical lattices of ultracold atoms},
  author = {Walsh, C. and S\'emon, P. and Sordi, G. and Tremblay, A.-M. S.},
  journal = {Phys. Rev. B},
  volume = {99},
  issue = {16},
  pages = {165151},
  numpages = {9},
  year = {2019},
  month = {Apr},
  publisher = {American Physical Society},
  doi = {10.1103/PhysRevB.99.165151},
  url = {https://link.aps.org/doi/10.1103/PhysRevB.99.165151}
}

@article{sordiPseudogapTemperatureWidom2012,
  title = {Pseudogap Temperature as a {{Widom}} Line in Doped {{Mott}} Insulators},
  author = {Sordi, G. and S{\'e}mon, P. and Haule, K. and Tremblay, A.-M. S.},
  year = 2012,
  month = jul,
  journal = {Scientific Reports},
  volume = {2},
  number = {1},
  pages = {547},
  issn = {2045-2322},
  doi = {10.1038/srep00547},
}

@article{varmaphenomenology1989,
  title = {Phenomenology of the normal state of Cu-O high-temperature superconductors},
  author = {Varma, C. M. and Littlewood, P. B. and Schmitt-Rink, S. and Abrahams, E. and Ruckenstein, A. E.},
  journal = {Phys. Rev. Lett.},
  volume = {63},
  issue = {18},
  pages = {1996--1999},
  numpages = {0},
  year = {1989},
  month = {Oct},
  publisher = {American Physical Society},
  doi = {10.1103/PhysRevLett.63.1996},
  url = {https://link.aps.org/doi/10.1103/PhysRevLett.63.1996}
}

@article{sachdevgapless1993,
  title = {Gapless spin-fluid ground state in a random quantum Heisenberg magnet},
  author = {Sachdev, Subir and Ye, Jinwu},
  journal = {Phys. Rev. Lett.},
  volume = {70},
  issue = {21},
  pages = {3339--3342},
  numpages = {0},
  year = {1993},
  month = {May},
  publisher = {American Physical Society},
  doi = {10.1103/PhysRevLett.70.3339},
  url = {https://link.aps.org/doi/10.1103/PhysRevLett.70.3339}
}

@article{parcolletNonFermi1999,
  title = {Non-Fermi-liquid regime of a doped Mott insulator},
  author = {Parcollet, Olivier and Georges, Antoine},
  journal = {Phys. Rev. B},
  volume = {59},
  issue = {8},
  pages = {5341--5360},
  numpages = {0},
  year = {1999},
  month = {Feb},
  publisher = {American Physical Society},
  doi = {10.1103/PhysRevB.59.5341},
  url = {https://link.aps.org/doi/10.1103/PhysRevB.59.5341}
}

@article{geoargesQuantumFluctuations2001,
  title = {Quantum fluctuations of a nearly critical Heisenberg spin glass},
  author = {Georges, A. and Parcollet, O. and Sachdev, S.},
  journal = {Phys. Rev. B},
  volume = {63},
  issue = {13},
  pages = {134406},
  numpages = {17},
  year = {2001},
  month = {Mar},
  publisher = {American Physical Society},
  doi = {10.1103/PhysRevB.63.134406},
  url = {https://link.aps.org/doi/10.1103/PhysRevB.63.134406}
}

@article{liStrangeMetal2024,
  title = {Strange Metal and Superconductor in the Two-Dimensional Yukawa-Sachdev-Ye-Kitaev Model},
  author = {Li, Chenyuan and Valentinis, Davide and Patel, Aavishkar A. and Guo, Haoyu and Schmalian, J\"org and Sachdev, Subir and Esterlis, Ilya},
  journal = {Phys. Rev. Lett.},
  volume = {133},
  issue = {18},
  pages = {186502},
  numpages = {8},
  year = {2024},
  month = {Oct},
  publisher = {American Physical Society},
  doi = {10.1103/PhysRevLett.133.186502},
  url = {https://link.aps.org/doi/10.1103/PhysRevLett.133.186502}
}

@article{hartnollStrangeMetallicHolography2010,
  title = {Towards Strange Metallic Holography},
  author = {Hartnoll, Sean A. and Polchinski, Joseph and Silverstein, Eva and Tong, David},
  year = 2010,
  month = apr,
  journal = {Journal of High Energy Physics},
  volume = {2010},
  number = {4},
  pages = {120},
  issn = {1029-8479},
  doi = {10.1007/JHEP04(2010)120},
}

@article{hartnollScalingTheory2015,
  title = {Scaling theory of the cuprate strange metals},
  author = {Hartnoll, Sean A. and Karch, Andreas},
  journal = {Phys. Rev. B},
  volume = {91},
  issue = {15},
  pages = {155126},
  numpages = {9},
  year = {2015},
  month = {Apr},
  publisher = {American Physical Society},
  doi = {10.1103/PhysRevB.91.155126},
  url = {https://link.aps.org/doi/10.1103/PhysRevB.91.155126}
}

@article{hartnollTheoryUniversalIncoherent2015,
  title = {Theory of Universal Incoherent Metallic Transport},
  author = {Hartnoll, Sean A.},
  year = 2015,
  month = jan,
  journal = {Nature Physics},
  volume = {11},
  number = {1},
  pages = {54--61},
  issn = {1745-2481},
  doi = {10.1038/nphys3174},
}

@article{Lohneysen_Rosch_Vojta_2007, title={Fermi-liquid instabilities at magnetic quantum phase transitions}, volume={79}, ISSN={0034-6861, 1539-0756}, DOI={10.1103/RevModPhys.79.1015}, number={3}, journal={Reviews of Modern Physics}, author={Löhneysen, Hilbert v. and Rosch, Achim and Vojta, Matthias and Wölfle, Peter}, year={2007}, month=aug, pages={1015–1075} }
